\documentclass[11pt,aps,prd,showpacs,nofootinbib,superscriptaddress,preprint,preprintnumbers]{revtex4}

\usepackage{natbib}
\usepackage{color}
\usepackage{graphicx}
\usepackage{amsmath}
\usepackage[caption=false]{subfig}
\usepackage[colorlinks=true,linkcolor=blue,citecolor=blue,urlcolor=black]{hyperref}

\newcommand{\be}{\begin{equation}}
\newcommand{\ee}{\end{equation}}
\newcommand{\bea}{\begin{eqnarray}}
\newcommand{\eea}{\end{eqnarray}}
\newcommand{\pptta}{$pp\to t \bar t \gamma+X\,$}
\newcommand{\TeV}{\,\rm{TeV}}
\newcommand{\GeV}{\,\rm{GeV}}

\usepackage{ulem,fancyvrb}

\usepackage{xcolor}
\usepackage{lipsum}
\usepackage{multirow}

\begin{document}
\title{Electroweak corrections to top quark pair production  in association with a hard photon at hadron colliders}
\author{Peng-Fei Duan}
\affiliation{City College, Kunming University of Science and Technology, \\
Kunming, Yunnan 650051, China}
\author{Yu Zhang \footnote{dayu@nju.edu.cn}}
\affiliation{School of Physics, Nanjing University, \\
 Nanjing, Jiangsu, 210093, China}
\affiliation{City College, Kunming University of Science and Technology, \\
Kunming, Yunnan 650051, China}
\author{Yong Wang}
\affiliation{Department of Modern Physics, University of Science and Technology of China, \\
Hefei, Anhui 230026, P.R.China}
\author{Mao Song}
\affiliation{School of Physics and Material Science, Anhui University, \\
Hefei, Anhui 230039, China}
\author{Gang Li}
\affiliation{School of Physics and Material Science, Anhui University, \\
Hefei, Anhui 230039, China}

\begin{abstract}
We present the next-to-leading order (NLO) electroweak (EW) corrections to the
top quark pair production associated with a hard photon at the current and future hadron colliders.
The dependence of the leading order (LO) and NLO EW corrected
cross sections on the photon transverse momentum cut are investigated. 
We also provide the LO and NLO  EW corrected distributions of the transverse momentum of final top quark 
and photon and the invariant mass of top quark pair and top-antitop-photon system. 
The results show that the NLO EW corrections are significant in high energy
regions due to the EW Sudakov effect.
\end{abstract}

\maketitle

%%%%%%%%%%%%
%%  section (intro)
%%%%%%%%%%%%

\section{Introduction \label{sec:intro}}

In the standard model (SM), the top quark is a very special particle. Because  its mass is much larger 
than any other SM elementary particles (except Higgs boson), the top quark is speculated to play a special role in electroweak 
symmetry breaking (EWSB). Since its unique properties have long been believed of potentially carrying 
important information to solve some of the paramount open questions in particle physics,
precise measurements of the cross sections and properties of top quark production channels are significant.
With the measurement of the cross section of top quark pair production in association with a hard photon, 
the strength of the electromagnetic coupling of the top quark and photon can be probed directly. 

Experimentally, measurements of the production rate of $t \bar t \gamma$ have been performed 
in $p\bar p$ collisions at the Tevatron by the CDF Collaboration at $\sqrt s = 1.96 \TeV$ \cite{Aaltonen:2011sp}
and in $pp$ collisions at the LHC by the ATLAS Collaboration at  $\sqrt s = 7 \TeV$ \cite{Aad:2015uwa}
and by the CMS Collaboration at  $\sqrt s = 8 \TeV$ \cite{CMS:1900ipz}.
From the theoretical point of view, the calculation of the cross section of $t \bar t \gamma$ production at hadron colliders
beyond the leading order (LO) used to be a very challenging problem.
The calculation of NLO QCD corrections to the production of 
$t \bar t$ pair and a hard photon at the Tevatron and the LHC have been performed in Refs.\cite{Duan:2009zza,PengFei:2011qg,
 Melnikov:2011ta}, which has a strong phenomenological motivation due to the large $K$-factor. 

With both the energy and luminosity increment in Run II of the LHC compared with Run I and future hadron colliders whose energy 
can be up to $100 \TeV$ \cite{benedikt,cepc_website} planed to be built, the need to increase the precision of the 
perturbative predictions becomes important and urgent. At fixed order, there are two ways \cite{Andersen:2014efa, 
Badger:2016bpw}: computing either the next-to-next-to-leading order (NNLO) QCD  or the 
NLO electroweak (EW) corrections, which are believed to be comparable numerically. 
Although the NLO EW correction is normally suppressed by the smallness of 
the coupling constant $\alpha$ 
and nominally subdominant with respect to
the QCD contributions, the NLO EW correction can become significant 
in the high-energy domain due to the
appearance of Sudakov logarithms \cite{Sudakov:1954sw, Fadin:1999bq, Ciafaloni:1998xg} that result from the
virtual exchange of soft or collinear massive weak gauge
boson. In this paper, we aim at the NLO EW corrections for
the $t\bar t\gamma$ production at the LHC and at future higher energy hadron colliders and present the results for the first time.

The rest of the paper is organized as follows: In section \ref{sec:calc}, we provide a general 
setup of our calculation. In section \ref{sec:results},
we present the numerical results and discussions for the LO and NLO EW corrected integrated and differential cross sections. 
Finally, a short summary is given in section \ref{sec:sum}.

%%%%%%%%%%%%
%%  section (Calculation setup)
%%%%%%%%%%%%

\section{Calculation setup \label{sec:calc}}

\begin{figure}[htbp]
\vspace{0.2cm}
\centering
\includegraphics[angle=0,width=4.8in,height=4.0in]{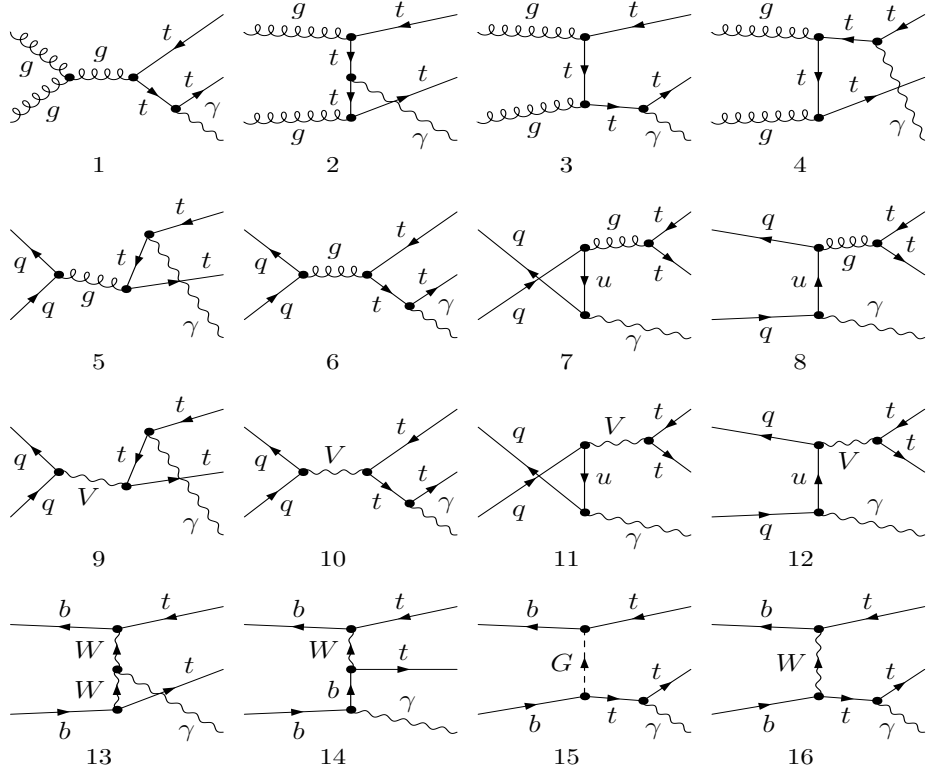}%
\caption{The representative Feynman diagrams at tree level, where $V=Z,\gamma$.
}
\label{fig:tree}
\end{figure}

For the process \pptta, at tree level the main partonic subprocesses are $gg\to t\bar t\gamma$ and $q \bar q \to t \bar t\gamma$,
where $q$ denotes the light quarks ($u, d, c, s, b$) if not otherwise stated. 
The corresponding representative Feynman diagrams are displayed in Fig.\ref{fig:tree}.
We can see that, the tree level amplitudes of the subprocess $gg\to t\bar t\gamma$ can be obtained from the 
Feynman diagrams in the first
line of Fig.\ref{fig:tree}(1-4) which are all at ${\cal O}(\alpha_s\alpha^{1/2})$,
while for the subprocess $q\bar q \to t \bar t \gamma$, the diagrams consist of two types: the gluon-mediated at 
${\cal O}(\alpha_s\alpha^{1/2})$ depicted in the second line of Fig.\ref{fig:tree}(5-8) and 
the $Z/\gamma$-mediated at ${\cal O}(\alpha^{3/2})$ depicted in the third line of Fig.\ref{fig:tree}(9-12).
Here, we assume that the CKM matrix is diagonal 
\footnote {For the bottom quark initial state $b\bar b$, another type will be exist, i.e., the $W$-exchanged 
at ${\cal O}(\alpha^{3/2})$, displayed in the last line of Fig.\ref{fig:tree}(13-16).}.
Then, the cross section will have different order of $\alpha_S$ and $\alpha$ and can be written as follows:
\bea
\sigma_{tree-level} &=& \sigma^{gg}(\alpha_S^2\alpha)+\sigma^{q\bar q}(\alpha_S^2\alpha)+
\sigma^{q\bar q}(\alpha^3)+\sigma^{b\bar b}(\alpha_S\alpha^2) \nonumber\\
&\equiv& \sigma_{LO,1}(\alpha_S^2\alpha)+\sigma_{LO,2}(\alpha_S\alpha^2)+\sigma_{LO,3}(\alpha^3)
\eea
This equation implicitly defines the leading, second-leading, and third-leading contributions in terms of the order of $\alpha_S$.
At the order of $\alpha_S\alpha^2$, the contributions from the gluon-mediated diagrams
interfacing with the $Z/\gamma$-mediated ones vanish owing to the color structure for the light quark initial states $q\bar q$, while 
the gluon-mediated diagrams interfacing with the $W$-exchanged ones survive for $b \bar b$ initial 
state \footnote {When the assumption of diagonal CKM matrix is relaxed, these contributions for $q\bar q\to t\bar t\gamma$ subprocess
can also be non-zero but are CKM-suppressed.}.
The first term $\sigma_{LO,1}(\alpha_S^2\alpha)$ is traditional contribution at LO and labelled as $\sigma_{LO}$ in this paper. 
%%%%%%%%%%%%%%%%%%%%%%%%%%%%%%
%%%%%%%%%%%%%%%%%%%%%%%%%%%%%%%

Analogously, at the one-loop level, i.e., NLO, one has
\bea
\Delta\sigma_{one-loop-level} \equiv \Delta\sigma_{NLO,1}(\alpha_S^3\alpha)
+\Delta\sigma_{NLO,2}(\alpha_S^2\alpha^2)+\Delta\sigma_{LO,3}(\alpha_S\alpha^3)
+\Delta\sigma_{LO,4}(\alpha^4)
\eea
The contributions from the first term $\Delta\sigma_{NLO,1}(\alpha_S^3\alpha)$ consist of the NLO QCD corrections which have been report in  
Refs.\cite{Duan:2009zza, Melnikov:2011ta}. The second term $\Delta\sigma_{NLO,2}(\alpha_S^2\alpha^2)$ 
labelled as $\Delta\sigma_{NLO,\,EW}$
will be calculated for the first time here, since it make up the NLO EW corrections. 
\footnote{The contributions at ${\cal O}(\alpha_S^2\alpha^2)$ can also come from the QCD corrections to 
$\sigma_{LO,2}(\alpha_S\alpha^2)$, which will be ignored here due to their tininess.}
Then, we define the NLO electroweak corrected
cross section as
\bea
\sigma_{NLO, EW} &=& \sigma_{LO}+\Delta\sigma_{NLO,\,EW} \nonumber \\
            &=& \sigma_{LO}\times(1+\delta_{NLO,EW})
\eea
where $\delta_{NLO,\,EW}\equiv \frac{\sigma_{NLO,\,EW}-\sigma_{LO}}{\sigma_{LO}}$
is the corresponding relative NLO EW corrections.

The calculation for the \pptta process is  performed by using the 't Hooft-Feynman gauge. 
The parent hadronic process \pptta is contributed by $gg \to t\bar{t} \gamma$ and $q \bar q \to t\bar{t} \gamma$ partonic processes, 
and the NLO EW corrections decompose two parts: the virtual and real emission correction. 
In the virtual correction, there exists ultraviolet (UV) and infrared (IR) divergences, which can be isolated by adopting  
the dimensional regularization (DR) scheme. After performing the renormalization procedure, the UV divergences can be removed 
through proper counter terms \cite{Ross:1973fp,Denner:1991kt} and a UV finite result will be obtained. 

In the calculation of the NLO EW corrections to the subprocess $gg\to t\bar t \gamma$, the photonic IR divergences originating from
exchange of virtual photon in loop can be cancelled with ones in the real photon emission correction 
from $gg\to t \bar t \gamma\gamma$ process. 
In order to extract the IR divergences in the virtual and real corrections, the two cutoff phase space slicing 
(TCPSS) method \cite{Harris:2001sx} is adopted. In addition, the dipole subtraction  (DS) method,
in which we transfer the dipole formulae in QCD provided in Refs.\cite{Catani:1996jh, Catani:1996vz,Catani:2002hc}
in a straightforward way to the case of dimensionally regularised photon emission, is used to verify the validity of the result. 

For the NLO EW corrections at ${\cal O}(\alpha_s^2\alpha^2)$ to the subprocess $q\bar q \to t \bar t\gamma$, 
the virtual contributions consist of two parts: 
the ${\cal O}(\alpha_s\alpha^{1/2})$ gluon-mediated tree-level amplitudes interfering with ${\cal O}(\alpha_s\alpha^{3/2})$ 
one-loop ones which may contain either photonic IR divergences or gluonic IR divergences 
and the ${\cal O}(\alpha^{3/2})$ $Z/\gamma$-mediated tree-level amplitudes interfering 
with ${\cal O}(\alpha_s^{2}\alpha^{1/2})$ one-loop ones which contain gluonic IR divergences. 
In order to cancel both the photonic and gluonic IR divergences in the virtual contributions, the corresponding 
real photon emission and real gluon emission corrections should be introduced at the same order of $\alpha_s^2\alpha^2$.
The contributions of real photon emission come from the Feynman diagrams at ${\cal O}(\alpha_s\alpha)$ 
of $q\bar q \to t\bar t \gamma \gamma$ subprocess,
and the real gluon emission corrections contribute by the Feynman diagrams at ${\cal O}(\alpha_s^{3/2}\alpha^{1/2})$ 
interfering with ones at ${\cal O}(\alpha_s^{1/2}\alpha^{3/2})$ of $q \bar q \to t \bar t \gamma g$ subprocess. 
Here we find that only the contributions from
the interference between the initial and final state gluon radiation diagrams are nonzero owing to the color structure. 
All the photonic and gluonic IR divergences in the virtual and real emission corrections are extract by the TCPSS method 
and the DS method are also applied to check. 

The FeynArts-3.7 package \cite{Hahn:2000kx} is applied to generate the Feynman diagrams automatically 
and the corresponding amplitudes are algebraically simplified by the FormCalc-7.2 program \cite{Hahn:1998yk}. 
In the calculation of one-loop Feynman amplitudes, we adopt the {\sc LoopTools-2.8} 
package \cite{Hahn:1998yk} for the numerical calculations of the scalar and tensor integrals, in which the $n$-point ($n\le 4$) 
tensor integrals are reduced to scalar integrals recursively by using Passarino-Veltman algorithm and the 5-point integrals are 
decomposed into 4-point integrals by using the method of Denner and Dittmaier \cite{Denner:2002ii}. In our previous 
work \cite{Yu:2013dxa,Chong:2014rea,Yu:2014cka, Yu:2015kbg, Zhang:2016kgc}, 
we addressed the numerical instability originating from the small 
Gram determinant ($detG$) and scalar one-loop 4-point integrals \cite{Boudjema:2009pw}.
In order to solve these instability problems  in the numerical calculations, we developed the {\sc LoopTools-2.8} package, 
which can automatically switch to the quadruple precision codes in the region of small Gram determinants, and checked the results with 
ones by using {\sc OneLoop} package \cite{vanHameren:2009dr}to verify the correctness of our codes. 

%%%%%%%%%%%%
%%  section (Numerical results)
%%%%%%%%%%%%
\section{Numerical results\label{sec:results}}
In our numerical evaluations we take the particles masses as follows:
\bea
M_W &=& 80.385\GeV, \qquad M_Z = 91.1876\GeV, \nonumber \\
M_H &=& 125\GeV, \qquad m_t = 173.5\GeV.
\eea
All widths are set equal to zero.
Our default EW scheme is the $\alpha(0)$ scheme, where we set:
\be
\alpha(0)=1/137.035999074
\ee

We adopt the MSTWlo2008\cite{Martin:2009iq} PDFs with the associated $\alpha_S(M_Z)$  for all NLO EW as
well as LO predictions, since we are chiefly interested in assessing effects of matrix-element
origin. We factorize and absorb initial state photonic  collinear singularities into the PDFs by using the
DIS factorization scheme. The renormalization ($\mu_R$) and the factorization ($\mu_F$) scales are set to be equal,
$\mu_R=\mu_F=m_t$. 

In order to exclude the inevitably IR divergence at tree level, we require 
the final state photon tagged hard with $p_{T}^{\gamma}>p_{T}^{\gamma,cut}$. 
If additional  photon  bremsstrahlung  is  present,  any
further phase-space cuts will only be applied to the visible photon with highest $p_T$, 
while the other is treated inclusively to ensure IR safety.

In table \ref{tab:ptcut}, we list the LO and NLO EW corrected cross sections and the corresponding relative NLO EW corrections 
to $t\bar t\gamma$ production at the 13 TeV LHC and 100 TeV proton-proton colliders for some typical values of 
$p_{T}^{\gamma, cut}$ separately. 
From this table we find that with the increment of $p_T^{\gamma,cut}$, the NLO EW correction becomes more 
significant due to the large EW Sudakov logarithms. For $p_T^{\gamma,cut}=1000~\GeV$, the NLO EW corrections 
are all almost $15\%$ for the 13 TeV LHC and 100 TeV hadron colliders. 
%%-----------------------------------------
\begin{table}
\center
\begin{tabular}{|c|c|ccccc|}
\hline
\multicolumn{2}{|c|}{$p^{\gamma, cut}_{T}$~[GeV]}   & 50 &  100  &  200 & 500 & 1000   \\
\hline
\multirow{2}{*}  
{$\sigma_{{\rm LO}}$} & $13\TeV$~[fb] & 851.4(3) &  356.1(2)  &  93.12(4) & 4.596(2) & 0.14778(4)   \\ \cline{2-7}
&  $100\TeV$~[pb] & 61.42(2) &  30.47(2)  &  10.616(6) & 1.1123(7) & 0.10661(7)   \\
\hline
\multirow{2}{*} 
{$\sigma_{{\rm NLO,\,EW}}$} &  $13\TeV$ ~[fb]& 835.4(4) &  348.4(3)  &  89.92(5) & 4.205(3) & 0.1259(4)   \\\cline{2-7}
&  $100\TeV$~[pb] & 60.04(3) &  29.69(3)  &  10.205(8) & 1.0158(9) & 0.09068(9)   \\
\hline
\multirow{2}{*} 
{$\delta_{{\rm NLO,\,EW}}$~[\%]} & $13\TeV$ & -1.9 &  -2.2  &  -3.4 & -8.5 & -14.8   \\\cline{2-7}
&  $100\TeV$ &   -2.2 &  -2.6  &  -3.9 & -8.7 & -14.9 \\
\hline
\end{tabular}
\caption{\small The LO and NLO EW corrected integrated cross sections and the corresponding relative NLO EW
corrections to the \pptta production at the $13~ {\rm TeV}$ LHC and 100 TeV proton-proton colliders for 
some typical values of $p^{\gamma,cut}_{T}$.}
\label{tab:ptcut}
\end{table}
%%%%%%%%%%%%%%%%%%%%%%

In the following, we turn to present results for kinematic distributions of final particles 
at the 13 TeV LHC with $p_{T}^{\gamma,cut} = 50 \GeV$. The relative NLO EW corrections to the differential cross
section $d\sigma/dx$ are defined as $\delta(x)=\left(\frac{d\sigma_{NLO,EW}}{dx}-\frac{\sigma_{LO}}{dx}\right)/\frac{\sigma_{LO}}{dx}$,
where $x$ denotes kinematic
observable. In the following, the considered observables contain the transverse momentum of the top quark ($p_T^t$) 
and the hard photon ($p_T^\gamma$) and the invariant mass of $t\bar t$ pair ($M_{t\bar t}$) and top pair association with 
a hard photon system ($M_{t\bar t\gamma}$).  

\begin{figure}[htbp]
\includegraphics[angle=0,width=3.2in,height=2.4in]{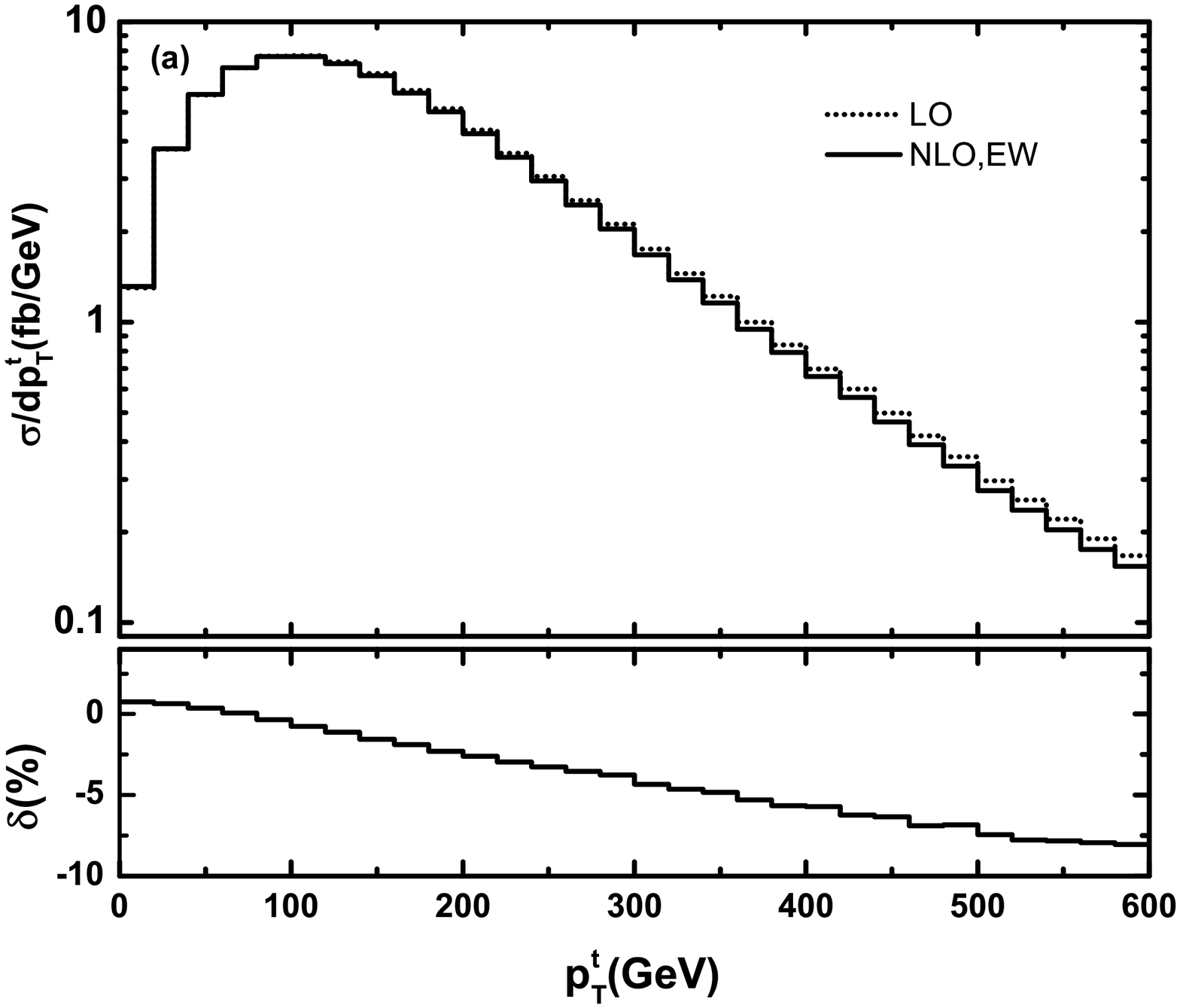}%
\hspace{0in}%
\includegraphics[angle=0,width=3.2in,height=2.4in]{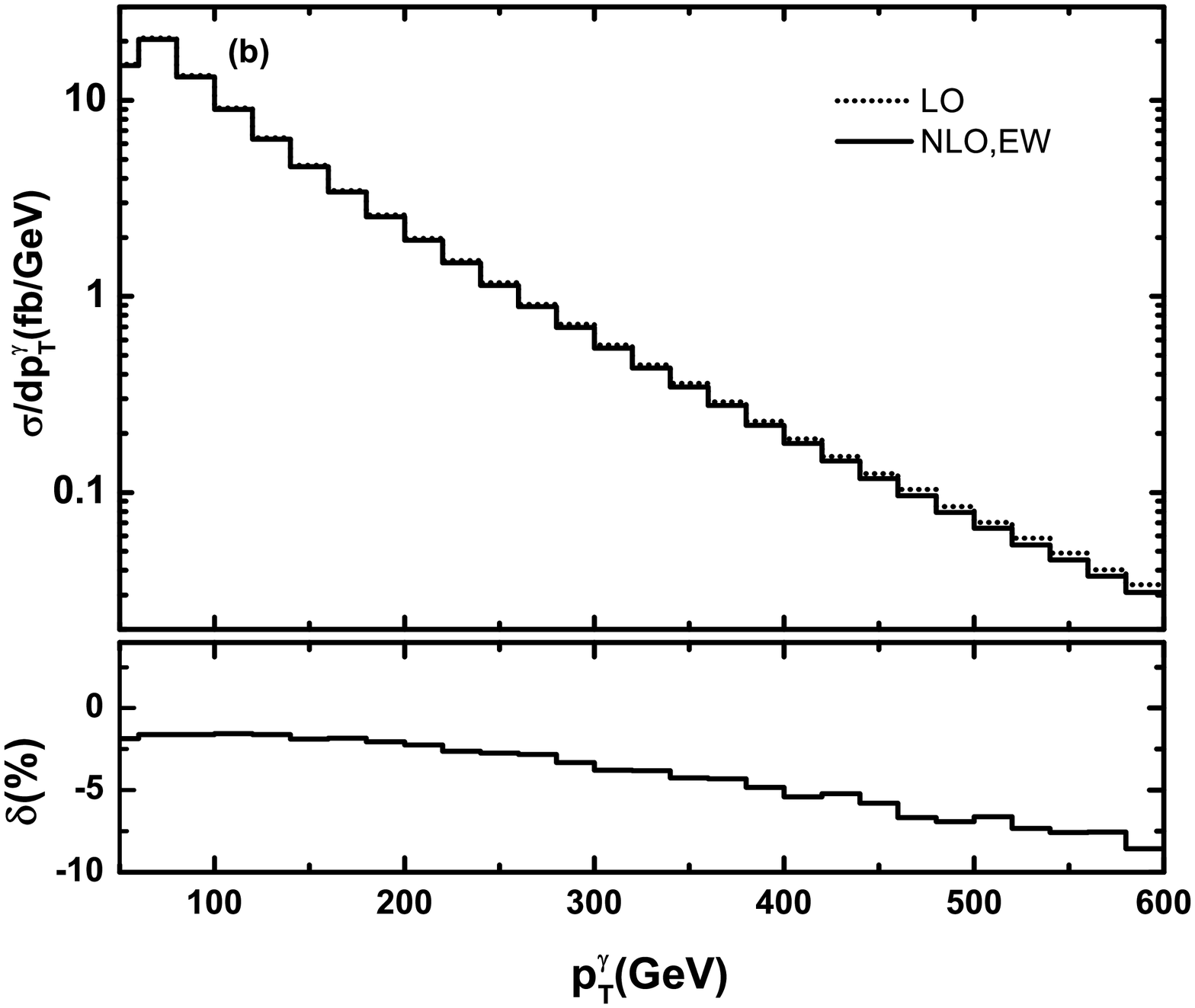}%
\hspace{0in}%
\caption{The LO, NLO EW corrected distributions and the relative NLO EW
corrections of \pptta process at 13 TeV
LHC with $p_T^{\gamma,cut}=50\GeV$ for $p_T^t$ (a) and $p_T^\gamma$ (b).  
}
\label{fig:pt}
\end{figure}

\begin{figure}[htbp]
\includegraphics[angle=0,width=3.2in,height=2.4in]{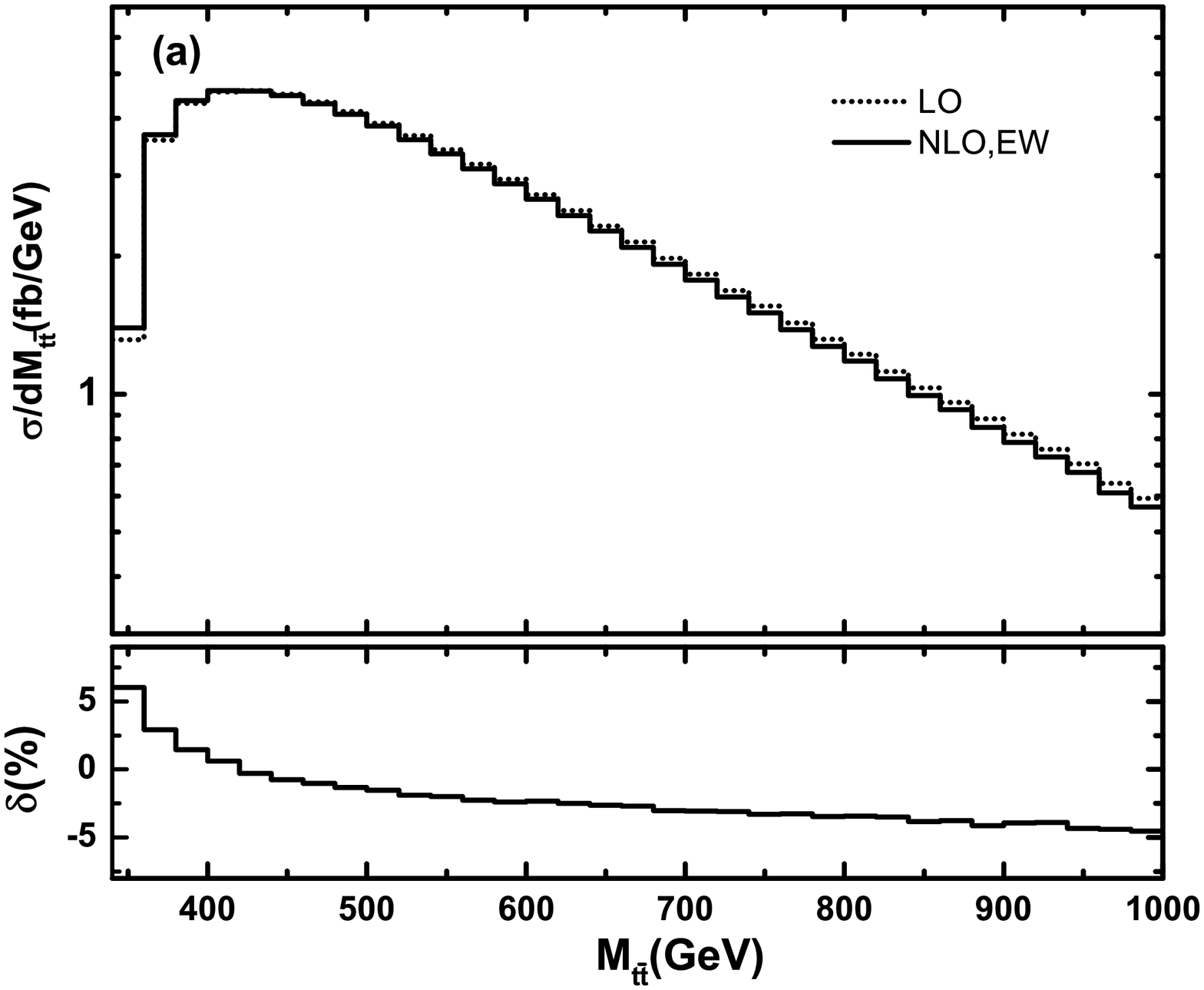}%
\hspace{0in}%
\includegraphics[angle=0,width=3.2in,height=2.4in]{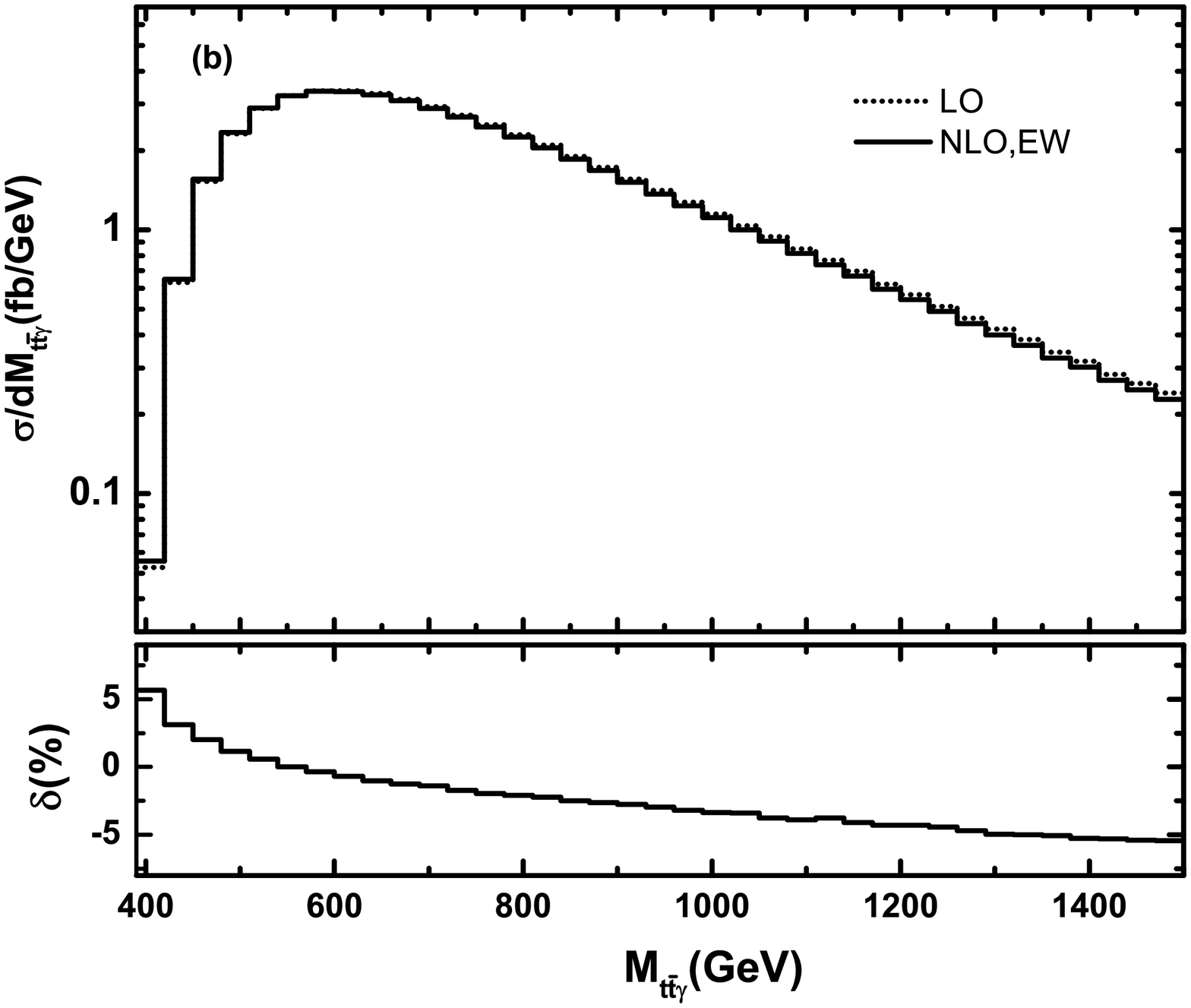}%
\hspace{0in}%
\caption{The LO, NLO EW corrected distributions and the relative NLO EW
corrections of \pptta process at 13 TeV
LHC with $p_T^{\gamma,cut}=50\GeV$ for $M_{t\bar t}$ (a) and $M_{t\bar t\gamma}$(b).  
}
\label{fig:im}
\end{figure}

In Fig.\ref{fig:pt}(a) and (b), we depict the LO and NLO EW corrected distributions for the 
transverse momentum of the top quark ($p_T^t$) and the hard photon ($p_T^\gamma$). 
The Fig.\ref{fig:im}(a) and (b) present the LO and NLO EW corrected invariant mass distributions of 
$t-\bar t$ ($M_{t\bar t}$) and $t-\bar t-\gamma$ ($M_{t\bar t\gamma}$)
system separately. The corresponding relative NLO EW corrections are also shown.
We can see that all the relative EW corrections to the considered four observables distributions mostly decrease with the increment 
of $p_T^t$, $p_T^\gamma$, $M_{t\bar t}$ and $M_{t\bar t\gamma}$ in the plotted region. The The LO $p_T^t$ distributions are 
enhanced by NLO EW corrections when $p_t^T<80\GeV$ and  suppressed in the rest plotted region, 
while the LO differential cross section of  $p_T^\gamma$ are always suppressed by
NLO EW corrections in the whole plotted region. The relative EW corrections to $M_{t\bar t}$ and $M_{t\bar t\gamma}$ turn 
to be negative when $M_{t\bar t\gamma}\ge 420 \GeV$ and $M_{t\bar t\gamma}\ge 570 \GeV$.
Due to the Sudakov effect, the absolute size of the NLO EW relative corrections continuously grow up with the increment of 
$p_T^t$, $p_T^\gamma$, $M_{t\bar t}$ and $M_{t\bar t\gamma}$  in the large region, particularly which can amount up to $-8.0\%$ 
at $p_T^{t}=600\GeV$, $-8.6\%$ at $p_T^{\gamma}=600\GeV$, $-4.5\%$ at $M_{t\bar t}=1000\GeV$ 
and $-5.4\%$ at $M_{t\bar t\gamma}=1500\GeV$.

%We have considered the following observables:
%the transverse momentum of the top quark ($p_T^t$) and the hard photon ($p_T^\gamma$) and the invariant mass of $t\bar t$ pair ($M_{t\bar t}$)

%%%%%%%%%%%%%%%%%%
%%%%%%%%%%%%%%%%%%%%%%%%%%%%%%

%%%%%%%%%%%%%%%%%%
%%%%%%%%%%%%%%%%%%%%%%%%%%%%%%

%%%%%%%%%%%%
%%  section (Summary)
%%%%%%%%%%%%

\section{Summary\label{sec:sum}}
In this work, we present the NLO EW corrections to the $t\bar t\gamma$ production at the 13 TeV LHC 
and future 100 TeV hadron colliders. Our results show that the NLO EW correction is significant in high energy 
region due to the EW Sudakov effect which can be most probably detected in the LHC experiments and future higher 
energy hadron colliders. For example, with the photon transverse momentum constraint of $p_T^{\gamma} > 1 \TeV$
on the hard photon, the NLO EW corrections to the total cross section of top quark pair production in association with a
photon at hadron colliders can reach about $-15\%$ with the collide energy $\sqrt s = 13 \TeV $ and 100 TeV. 
We also investigate the NLO EW effects on the kinematic distributions of $p_T^t$, $p_T^\gamma$, $M_{t\bar t}$ and $M_{t\bar t\gamma}$,
and find that the NLO EW corrections become significant in high energy regions.

\section*{Acknowledgements}
We would like to thank Zuowei Liu for reading and correcting the grammar and spelling mistakes.
This work was supported in part by the National Natural Science Foundation of China (Grant No.11405076, No.11347101, No.11305001,
No.11205003), the Applied Basic Research Programs of Yunan Provincial Science and Technology Department (Grant No.2016FB008), 
and the Startup Foundation for Doctors of Kunming University of Science and Technology(No.KKSY201356060).

\end{document}